%% file: submission.tex
\documentclass[runningheads]{llncs}
\usepackage[T1]{fontenc}
\usepackage{graphicx}
\usepackage{hyperref}

\usepackage{color}

\usepackage{booktabs}
\usepackage{multirow}
\usepackage[dvipsnames]{xcolor}
\usepackage{listings}
\newcommand\YAMLcolonstyle{\color{red}\mdseries}
\newcommand\YAMLkeystyle{\color{black}\bfseries}
\newcommand\YAMLvaluestyle{\color{blue}\mdseries}

\makeatletter

\newcommand\language@yaml{yaml}

\expandafter\expandafter\expandafter\lstdefinelanguage
\expandafter{\language@yaml}
{
  keywords={true,false,null,y,n},
  keywordstyle=\color{darkgray}\bfseries,
  basicstyle=\YAMLkeystyle\ttfamily,  % assuming a key comes first
  sensitive=false,
  comment=[l]{\#},
  morecomment=[s]{/*}{*/},
  commentstyle=\color{purple}\ttfamily,
  stringstyle=\YAMLvaluestyle\ttfamily,
  moredelim=[l][\color{orange}]{\&},
  moredelim=[l][\color{magenta}]{*},
  moredelim=**[il][\YAMLcolonstyle{:}\YAMLvaluestyle]{:}, % switch to value style at :
  morestring=[b]',
  morestring=[b]",
  literate =    {---}{{\ProcessThreeDashes}}3
                {>}{{\textcolor{red}\textgreater}}1     
                {|}{{\textcolor{red}\textbar}}1 
                {\ -\ }{{\mdseries\ -\ }}3,
}

\lst@AddToHook{EveryLine}{\ifx\lst@language\language@yaml\YAMLkeystyle\fi}
\makeatother

\newcommand\ProcessThreeDashes{\llap{\color{cyan}\mdseries-{-}-}}
\usepackage{algorithm}
\usepackage[]{algpseudocodex}
\usepackage{siunitx}
\usepackage{amsfonts}
\usepackage{amsmath}
\usepackage{orcidlink}

\begin{document}

\title{It's all about PR}
\subtitle{Smart Benchmarking AI Accelerators using \\\underline{P}erformance \underline{R}epresentatives}

\author{Alexander Louis-Ferdinand Jung\orcidlink{0000-0001-5702-5768} \and
Jannik Steinmetz\orcidlink{0009-0003-7193-5620} \and
Jonathan Gietz\orcidlink{0009-0000-7843-1781} \and
Konstantin Lübeck\orcidlink{0000-0002-2701-5881} \and
Oliver Bringmann\orcidlink{0000-0002-1615-507X}
}

\authorrunning{A. L.-F. Jung et al.}

\institute{Embedded Systems, University of Tübingen, Tübingen, Germany\\
}

\maketitle              % typeset the header of the contribution
\vspace{-6mm}
\begin{abstract}
\input{00_abstract}
\end{abstract}

\input{01_introduction}
\input{02_related_work}
\input{03_smart_benchmarking}
\input{04_evaluation}\vspace{-12mm}
\input{99_conclusion}

\begin{credits}
\subsubsection{\ackname} This work has been funded by the German Federal Ministry of Education and Research (BMBF) under grant number 16ES0876 (GENIAL!).

\subsubsection{\discintname} The authors have no competing interests to declare that are relevant to the content of this article.
% It is now necessary to declare any competing interests or to specifically
% state that the authors have no competing interests. Please place the
% statement with a bold run-in heading in small font size beneath the
% (optional) acknowledgments\footnote{If EquinOCS, our proceedings submission
% system, is used, then the disclaimer can be provided directly in the system.},
% for example: The authors have no competing interests to declare that are
% relevant to the content of this article. Or: Author A has received research
% grants from Company W. Author B has received a speaker honorarium from
% Company X and owns stock in Company Y. Author C is a member of committee Z.
\end{credits}
\bibliographystyle{splncs04}
\bibliography{references}

\end{document}

%% file: 00_abstract.tex
\label{sec:abstract}
Statistical models are widely used to estimate the performance of commercial off-the-shelf (COTS) AI hardware accelerators. However, training of statistical performance models often requires vast amounts of data, leading to a significant time investment and can be difficult in case of limited hardware availability.

To alleviate this problem, we propose a novel performance modeling methodology that significantly reduces the number of training samples while maintaining good accuracy. Our approach leverages knowledge of the target hardware architecture and initial parameter sweeps to identify a set of \textit{Performance Representatives} (PR) for deep neural network (DNN) layers. These PRs are then used for benchmarking, building a statistical performance model, and making estimations. This targeted approach drastically reduces the number of training samples needed, opposed to random sampling, to achieve a better estimation accuracy.

We achieve a Mean Absolute Percentage Error (MAPE) of as low as \SI{0.02}{\percent} for single-layer estimations and \SI{0.68}{\percent} for whole DNN estimations with less than \num{10000} training samples. The results demonstrate the superiority of our method for single-layer estimations compared to models trained with randomly sampled datasets of the same size.

\keywords{Performance Estimation \and Benchmarking \and AI Hardware Accelerators \and Statistical Modelling \and Deep Neural Networks}

%% file: 01_introduction.tex
\section{Introduction}
\label{sec:introduction}
Deep Neural Networks (DNNs) are being used in various tasks ranging from time-series and image classification to natural language processing. As these models become part of more and more AI-assisted edge devices, there is an abundance of commercial off-the-shelf AI accelerator hardware. Determining which accelerator to use for a specific AI application is difficult, as real hardware might not yet be available and simulations tend to be very time-consuming. Therefore, it is important to be able to estimate the performance of a DNN mapped onto an accelerator platform.

Another area where performance estimation models are needed is Neural Architecture Search (NAS). Especially for hardware-aware NAS targeting embedded platforms, a trade-off between a high DNN accuracy and the execution time has to be considered. However, measuring all possible DNN candidates can be extremely time-consuming. Therefore, an efficient performance estimator is needed to accelerate the optimization loop. \cite{Gerum2022}

Statistical estimation models have been shown to have high accuracy for execution time estimation if a sufficient amount of training samples is obtainable \cite{Wess2021,Bouzidi2021,Zhang2021}. Analytical models, on the other hand, require detailed knowledge about the architecture to yield accurate estimations \cite{Parashar2019,Lubeck2022}.

Vendors of commercial off-the-shelf (COTS) accelerators often only provide a slow functional and timing simulator. Gathering a sufficiently large amount of training samples for statistical estimation models with these simulators is infeasible. Another problem with COTS platforms is, that often only general specifications such as the available functional units, and the memory bandwidth are publicly known. An accurate analytical performance model cannot be created from such limited information. But these known hardware characteristics can indicate interesting behavior with specific DNN layer parameters, e.g., which tile-sizes yield a good utilization.

\begin{figure}[!t]
\centering
\includegraphics[width=0.9\textwidth]{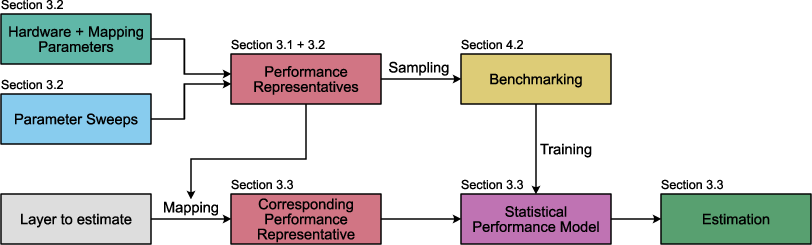}
\caption{The proposed performance modeling methodology and benchmarking strategy.}
\label{fig:general_flow}
\end{figure}

This paper proposes a performance modeling methodology paired with a benchmarking strategy to select significant data points to reduce measurement runs and consequently measurement time. Depending on the amount of knowledge that is available for a platform, the known hardware characteristics can be used to find DNN layer parameters that are representative of the execution time behavior. If there is not much information about a platform, these \textit{Performance Representatives} need to be determined empirically in an initial benchmark phase. An overview of the proposed approach is depicted in Fig.~\ref{fig:general_flow}.

This paper makes the following contributions: (1) We describe a methodology to determine the Performance Representatives of a specific hardware platform. (2) We then use these Performance Representatives to carry out benchmarks on the hardware platform under consideration and then use these representative data points to train statistical models which can then be used to estimate the execution time of single-layers and whole DNNs. (3) We also show how the amount of training samples impacts the accuracy when using our approach compared to random sampling the same amount of data points.

%% file: 02_related_work.tex
\section{Related Work}
\label{sec:related-work}
In ANNETTE \cite{Wess2021} they use micro-kernel and multi-layer benchmarks of DNN layers to create machine learning estimation models. These Random Forest models are then combined with the well-known Roofline model \cite{Williams2009} to better capture the regular execution time behavior of these platforms. They evaluate their mixed models on two hardware platforms (an FPGA and an ASIC accelerator) and achieve a mean absolute percentage error (MAPE) of as low as \SI{12.71}{\percent} for single-layer estimations and \SI{3.47}{\percent} for whole networks. Per platform and per layer \num{35000} measurements were conducted.

Bouzidi et al. \cite{Bouzidi2021} compare five different machine learning methods for performance estimation on the NVIDIA Jetson AGX Xavier and TX2 GPUs. They measured around \num{200000} whole DNN inferences, by using the execution times of the executed GPU kernels reported by NVIDIA's profiling tools. The MAPE of the best estimation model ranges between \SI{7.67}{\percent} and \SI{14.73}{\percent}.

The authors of Blackthorn \cite{Lechner2021} specifically exploit the very regular execution time behavior of the NVIDIA Jetson TX2 and the Jetson Nano to create their performance estimation models. They define a set of step-wise functions and reduce this set of functions gradually by measuring more and more data points. Each function that cannot explain the new data point is removed from the function set, resulting in one final step-wise function that is then used for estimating the performance. 
This requires around \num{15000} measurements for the Convolution layer, resulting in a Root-Mean-Square-Percentage-Error (RMSPE) of \SI{7.57}{\percent} for Convolution layers and an error between \SI{0.45}{\percent} and \SI{6.68}{\percent} for complete DNN execution time estimation. 

In nn-Meter \cite{Zhang2021} DNN execution kernels are benchmarked, which contain multiple fused layers and are used for estimating the execution time of various DNNs by summing the estimated times for the contained kernels. They obtain up to \num{39968} benchmark points for e.g. the Convolution execution kernel. These samples are then used for training Random Forest models, which yield a RMSPE between \SI{1.35}{\percent} and \SI{22.25}{\percent} depending on the DNN and hardware platform.

The presented estimators require a vast amount of training samples to achieve the reported estimation accuracies. Additionally, the usage of step-wise functions by Blackthorn has some drawbacks. Firstly, a function pattern that describes the execution time behavior needs to be defined. 
Secondly, the step-wise function describing the execution time behavior must be covered by the parameter space of the function set. This is a prerequisite so that the approach finds a matching function. We address these shortcomings by, on the one hand, reducing the number of training samples, and on the other hand, determining the step width algorithmically instead of choosing from a set of pre-defined functions for an accurate execution time estimation.

%% file: 03_smart_benchmarking.tex
\section{Smart Benchmarking and Modeling}
\label{sec:smart-benchmarking}
\begin{figure}[!t]
\centering
\includegraphics[width=0.9\textwidth]{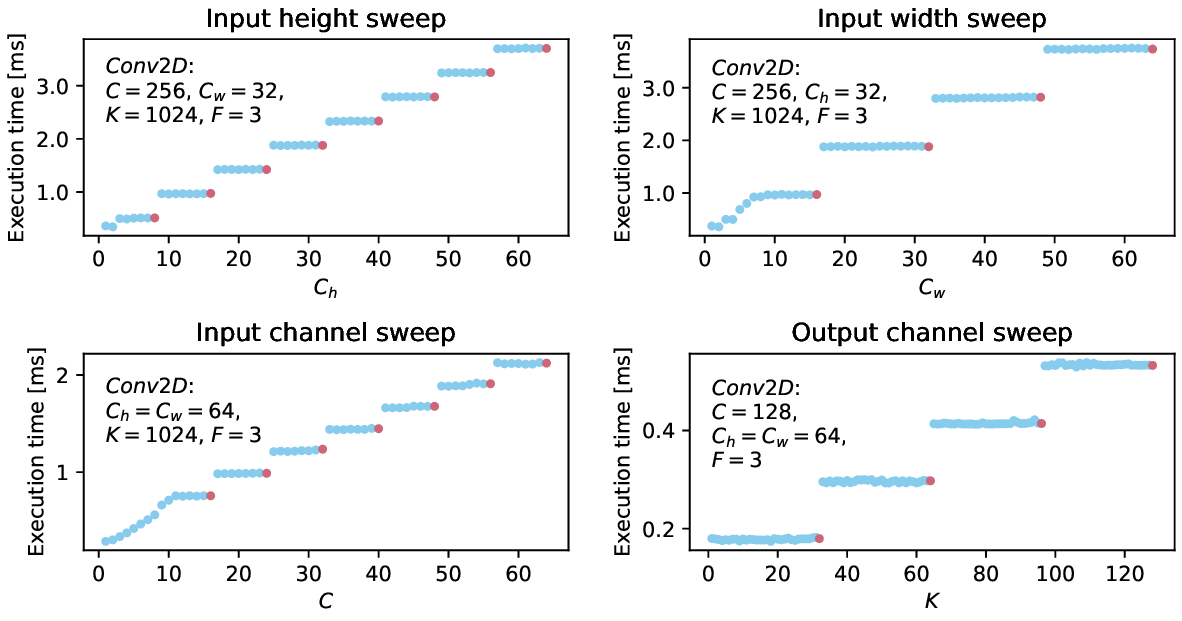}
\vspace{-3mm}
\caption{Step-wise execution time behavior of the 2D-Convolution observed for different parameter sweeps, run on the NVIDIA Jetson AGX Xavier GPU. Performance Representatives are highlighted in red (last point on step). The parameters of the 2D-Convolution are described in Eq. \ref{eq:conv2d} and its explanation.
}
\vspace{-3mm}
\label{fig:agx_sweep}
\end{figure}

The execution time behavior of an AI accelerator platform often exhibits a very regular pattern. This is due to how AI accelerators compute the layers of a DNN. For example, the UltraTrail accelerator \cite{Bernardo2020} always processes $8 \times 8$ output and input channels in one activation of the multiply accumulate (MAC) array. Therefore, the execution time behavior takes the shape of a step-wise function while increasing only the input channel parameter or the output channel parameter of the 1D-Convolution. We call these types of benchmarks \textit{parameter sweeps}. But also other hardware platforms, like the NVIDIA Jetson AGX Xavier GPU exhibit a similar behavior, as can be seen in Fig. \ref{fig:agx_sweep}.

This also means, that many of the data points in fine-grained measurements do not contain new information about the execution time behavior of the platform. The aim of this work is to reduce the number of measured data points for building statistical performance models, while maintaining good accuracy when estimating the execution time of a single layer as well as of whole DNNs.

\subsection{Performance Representatives (PR)}

In our work, we combine the findings of \cite{Lechner2021} and \cite{Wess2021}. Exploiting the regular execution time behavior to guide the selection of benchmark points, consequently reducing the number of points that need to be measured. These measurements are then used to train statistical models for execution time estimation. The idea is to only measure one representative data point for a step. We call this a \textbf{Performance Representative (PR)}. We algorithmically determine the last point of a step as our PR, as these are generally the points with optimal utilization of the AI accelerator architecture.

\subsection{Hardware and Mapping Parameters}
\begin{figure}[htbp]
\centering
\includegraphics[width=0.75\textwidth]{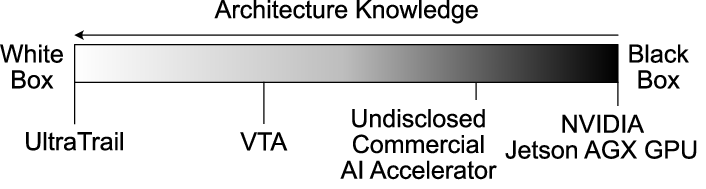}
\vspace{-3mm}
\caption{Accelerator classification on the white-box to black-box spectrum.}
\label{fig:accelerator-spectrum}
\end{figure}

How the PRs of the execution time behavior can be derived heavily depends on how much knowledge about the accelerator architecture is available. We classify the amount of available architecture knowledge for the accelerators used in this work ranging from white-box to black-box. This spectrum is depicted in Fig.~\ref{fig:accelerator-spectrum}.%
\vspace{-3mm}

\vspace{-3mm}

\textbf{White-box accelerators} are architectures where all the needed information about how the accelerator works and DNN layers are mapped is publicly available so that the execution time behavior of the architecture can be derived, for instance from the number of processing elements (PEs) in each dimension and how the layer is unrolled onto the architecture.

At the other end of the spectrum are the \textbf{black-box accelerators}. For those architectures, almost no information is available and therefore, an execution time behavior cannot be determined without measurements.

When some information about the AI accelerator architecture is available, but not enough to accurately determine the execution time behavior, we speak of \textbf{gray-box accelerators}. For this category, it is necessary to use the available information to guide the selection of benchmarks that will help to either confirm the expected PRs or to determine the execution time behavior from which the PRs will be identified. Depending on the category, the accelerator belongs to, the process of finding the PRs differs:\\

As stated above, for \textbf{white-box accelerators}, all the necessary information is available, and the PRs can be determined from it. To illustrate this, we use the UltraTrail accelerator by Bernardo et al. \cite{Bernardo2020} as an example. In the standard configuration, it only supports 1D-Convolutions:%
\vspace{-1mm}
\begin{equation}
    Conv1D(C, C_w, K, F, s, pad)
\end{equation}%
\vspace{-6mm}%

\noindent with input channels $C$, input feature width $C_w$, output channels $K$, kernel size $F$, stride $s$ and padding $pad$. In \cite{Bernardo2020}, the UltraTrail accelerator is described in detail so that it is possible to determine the hardware and mapping characteristics:

\begin{lstlisting}[language=yaml]
operation: Conv1D
operation_params: [C, C_w, K, F, s, pad] 
dims: [8, 8]
mapping: [C, K]
\end{lstlisting}
\noindent This hardware and mapping description specifies that the parameters $C$ and $K$ of the Conv1D layer are mapped and unrolled on the two spatial dimensions of the UltraTrail accelerator, each with a size of eight. From this information, the PRs can be determined as:%
\vspace{-2mm}
\begin{equation}
    Conv1D_R(x_C \cdot 8, C_w, x_K \cdot 8, F, s, pad)
\end{equation}%
\vspace{-6mm}

\noindent with $x_C, x_K \in \lbrace 1, 2, ..., 7 \rbrace$. The upper bounds of $x_C$ and $x_K$ stem from the upper bounds of the parameters $C$ and $K$ as described in \cite{Bernardo2020}.

As long as the AI accelerator architecture and mapping can be described like this, the PRs can be determined by setting the values of the unrolled parameters to a multiple of the respective dimensions.\\

In the category of \textbf{black-box accelerators}, we have AI accelerator architectures, for which there is almost no information available. Therefore, the PRs can only be determined from measurements. To illustrate this, we look at the NVIDIA Jetson AGX Xavier GPU. E.g. we do not know how the TensorRT\footnote{\url{https://developer.nvidia.com/tensorrt}, last accessed 03/26/2024} backend maps the Convolution layer%
\vspace{-2mm}
\begin{equation}
    Conv2D(C, C_h, C_w, K, F_h, F_w, s, pad)
    \label{eq:conv2d}
\end{equation}%
\vspace{-6mm}

\noindent with input channels $C$, input feature height $C_h$ and width $C_w$, output channels $K$, kernel height $F_h$ and width $F_w$, stride $s$ and padding $pad$ onto the Jetson AGX Xavier GPU. Therefore, we need to perform initial sweep benchmarks over the parameters $ P = \lbrace C, C_h, C_w, K, F_h=F_w \rbrace$, that influence the execution time behavior the most. We only consider quadratic kernel sizes, as those are most commonly used in state-of-the-art DNNs.

We use the procedure described in Algorithm~\ref{algo:representatives} to determine the step width $w_p \in W$ of each parameter $p \in P$ from the performed sweeps. In general, the PRs of the Conv2D layer can be determined from the step width $w_p$ as follows:%
\vspace{-2mm}
\begin{equation}
\begin{split}
    Conv2D_R(&x_C \cdot w_C, x_{C_h} \cdot w_{C_h}, x_{C_w} \cdot w_{C_w}, \\
    &x_K \cdot w_K, x_{F_h} \cdot w_{F_h}, x_{F_w} \cdot w_{F_w}, s, pad)
\end{split}
\end{equation}%
\vspace{-4mm}

\noindent where $x_p \in \mathbb{N}$ are the factors which determine each step (as an integer multiple of the step width). $x_p$ might have an upper bound depending on the capabilities of the AI accelerator. If Algorithm~\ref{algo:representatives} detects a linear influence of a parameter, it sets the step width $ w_p = 1$. The described procedure works analogously for other layer types as well.

Although we performed parameter sweeps on the NVIDIA Jetson AGX GPU for Fully-Connected, Pooling, and activation layers as well, only the following layers exhibited a non-linear execution time behavior from which the following PRs could be determined:%
\vspace{-2mm}
\begin{itemize}
\setlength\itemsep{0.3em}
    \item $Conv2D_R(x_C \cdot 8, x_{C_h} \cdot 8, x_{C_w} \cdot 16, x_K \cdot 32, F_h, F_w, s, pad)$ with $F_h = F_w \geq 3$
    \item $PointwiseConv2D_R(x_C \cdot 8, x_{C_h} \cdot 4, x_{C_w} \cdot 4, x_K \cdot 32, F_h=F_w=1, s, pad)$
    \item $DepthwiseConv2D_R(x_C \cdot 8, C_h, C_w, K=1, F_h, F_w)$ with input channels $C$, input height and width $C_h$ and $C_w$, depthwise multiplier $K$ and $F_h = F_w \geq 3$
\end{itemize}
\begin{algorithm}[ht]
\caption{Determine PRs from DNN layer parameter sweeps.}
\label{algo:representatives}
\scriptsize{
\begin{algorithmic}[1]
    \Require $S: P \to (\vec{x}$, $\vec{y}) \gets$ Sweep measurements over relevant layer parameters $P$, $threshold_{linear}: \mathbb{R}^+$
    \Function{TestLinearBehavior}{$\vec{x}$, $\vec{y}$, $threshold_{linear}$}
        \State $y_{min} \gets min(\vec{y})$, $y_{max} \gets max(\vec{y})$
        \Comment{Get min and max of the values in $\vec{y}$}
        \State $x_{min} \gets min(\vec{x})$, $x_{max} \gets max(\vec{x})$
        \Comment{Get min and max of the values in $\vec{x}$}
        \State $slope_{avg} \gets \frac{y_{max} - y_{min}}{x_{max} - x_{min}}$
        \Comment{Compute average slope over the sweep measurements}
        \State $\hat{\vec{y}} \gets slope_{avg} \cdot \vec{x} + x_{min}$
        \Comment{Linear erstimation of the execution time behavior}
        \State $rmse \gets \mathrm{RMSE}(\vec{y}, \hat{\vec{y}})$
        \Comment{RMSE between measured and linear estimation}
        
        \If{$rmse < threshold_{linear}$}
            \State \Return True
            \Comment{Parameter has linear influence on execution time}
        \Else
            \State \Return False
        \EndIf
    \EndFunction

    \Function{ExecutionTimeDelta}{$\vec{y}$}
        \State $\vec{deltas} \gets [\ ]$
        \ForAll{$y_i, y_{i+1} \in \vec{y}$}
        \Comment{Pairs of consecutive data points}
            \State $\vec{deltas}.append(y_{i+1} - y_i$)
        \EndFor
        \State \Return $\vec{deltas}$
    \EndFunction

    \State $W: P \to \mathbb{N}$
    \Comment{Step widths with which the parameters influence the execution time}
    \ForAll{$p, (\vec{x}, \vec{y}) \in S$}
    \Comment{Layer parameter $p$ with corresponding sweep over values $\vec{x}$ and measured execution time $\vec{y}$}
        \If{\Call{TestLinearBehavior}{$\vec{x}$, $\vec{y}$, $threshold_{linear}$}}
            \State $W[p] \gets 1$
        \Else
            \State $\vec{deltas} \gets$ \Call{ExecutionTimeDelta}{$\vec{y}$}
            \State $\vec{indices} \gets$ \Call{FindPeaks}{$\vec{deltas}$}
            \Comment{E.g. \texttt{scipy.signal.find\_peaks()}}
            \State $w_p \gets$ \Call{PeakDistance}{$\vec{indices}$}
            \Comment{Distance between peaks is the step width}
            \State $W[p] \gets w_p$
        \EndIf
    \EndFor
    \State \Return $W$
\end{algorithmic}
}
\end{algorithm}
\textbf{Gray-box accelerators} is the category that falls in between the two extremes, for which we provide two example AI accelerator architectures. The first one is an undisclosed commercial AI accelerator, for which the vendor provided us with a timing simulator under a non-disclosure agreement. We determined the execution time behavior by running initial sweep benchmarks and used Algorithm~\ref{algo:representatives} to find the step widths. We found a non-linear execution time behavior for the following layers, resulting in the PRs:
\begin{itemize}
\setlength\itemsep{0.3em}
    \item $Conv2D_R(C, x_{C_h} \cdot 8, x_{C_w} \cdot 8, x_K \cdot 16, F_h, F_w, s, pad)$
    \item $DepthwiseConv2D_R(C, x_{C_h} \cdot 8, x_{C_w} \cdot 8, K=1, F_h, F_w)$ with input channels $C$, input height $C_h$, input width $C_w$, depthwise multiplier $K$ and $F_h = F_w \geq 3$
    \item $FullyConnected_R(1, in, x_{out} \cdot 64)$ with inputs $in$ and outputs $out$
    \item $AveragePool2D_R(C, x_{C_h} \cdot 8, x_{C_w} \cdot 8, F)$ with input channels $C$, input height, $C_h$, input width $C_w$, and pooling kernel size $F$
\end{itemize}

From the information that is available to us, we additionally know that this AI accelerator contains two separate functional units (FU). One optimized for big Convolution and Fully-Connected operations and one for different activation functions, Max- and Average-Pooling, and smaller Convolutions. These FUs can execute consecutive layers in a DNN in an overlapping fashion. We therefore constructed multi-layer benchmarks, similar to those in \cite{Wess2021}, to verify this specific behavior. This information is used to improve multi-layer and whole DNN estimations.

The second example is the Versatile Tensor Accelerator (VTA) \cite{Moreau2019} that is part of the Apache TVM project\footnote{\url{https://tvm.apache.org/}, last accessed 03/26/2024}. In this case, an expert could theoretically retrieve all the required information from the open-source Chisel implementation. But most of the time, working with only semi-detailed information is necessary due to practical limitations. To illustrate our methodology, we assume that we only know the following about VTA: The supported operations are 2D-Convolution, general matrix multiplication (GeMM), Fully-Connected, and element-wise matrix operations. In the used configuration of the VTA hardware, the GeMM core can compute a $(1, 16) \times (16, 16)$ matrix-matrix multiplication per cycle. Furthermore, the documentation suggests that for a multiplication of matrices $A$ and $B$, the columns of $A$, as well as the columns of $B$ have to be an integer multiple of \num{16} or padding needs to be applied. Analogously, for the 2D-Convolution, the parameters input channels $C$ and output channels $K$ have to be integer multiples of 16 or again padding is necessary. To confirm these PRs, we performed sweeps for the GeMM and the 2D-Convolution operation, which revealed that the PRs are indeed: \vspace{-3mm}
\begin{equation}
    Conv2D_R(x_C \cdot 16, C_h, C_w, x_K \cdot 16, F_h, F_w, s, pad) \text{\quad and}
\end{equation}%

\vspace{-4.5mm}
\begin{equation}
    FullyConnected_R(1, x_{in} \cdot 16, x_{out} \cdot 16)
\end{equation}% 
\vspace{-8mm}

\subsection{Performance Estimation using Performance Representatives}
As our goal is to systematically reduce the number of benchmark points for training an estimator while still maintaining a good estimation accuracy, we use Random Forest Regression to build the estimator as they have been shown to be very accurate for DNN execution time estimation as described in \cite{Wess2021,Zhang2021} but other machine learning methods would also have been applicable \cite{Bouzidi2021}.

After determining the PRs of an accelerator, we perform measurements using only data points from the set of PRs. Most of the time, even this constrained set of layer configurations is too large to be benchmarked completely. E.g. the set of PRs of the $Conv1D$ operation supported by UltraTrail, still contains \num{1493520} possible configurations (the complete parameter space contains \num{95585280} data points). Consequently, we need to sample from the set of PRs. In this study, we explore the accuracy of our approach, without aiming for a specific target accuracy of the estimation. However, the benchmark sampling and model training could easily be used in an optimization loop until the desired estimation accuracy is achieved.

A relevant aspect to consider when using Random Forests is that this type of model cannot extrapolate, but only estimate in the range of values it has seen during training. Therefore, it is important to know for which DNN and, consequently, for which layer parameters a model should be able to make estimations and ensure that the range of benchmark points covers the area of interest.

\subsubsection{Single-Layer Execution Time Estimation}
When it comes to making an estimation for a layer, we first map the layer configuration in question to the corresponding PR by selecting the next larger integer multiple of the step width $w_p$ for every parameter. For the $Conv2D$ operation, this looks like this:
\begin{equation}
\label{eq:pr_mapping}
\begin{split}
    Conv2D(&C, C_h, C_w, K, F_h, F_w, s, pad) \rightarrow \\    
    Conv2D_R(&x_C \cdot w_C, x_{C_h} \cdot w_{C_h}, x_{C_w} \cdot w_{C_w}, \\
    &x_K \cdot w_K, x_{F_h} \cdot w_{F_h}, x_{F_w} \cdot w_{F_w}, s, pad) \\
\end{split}
\end{equation}
\begin{equation}
    \text{with\quad} x_p = \biggl \lceil \frac{p}{w_p} \biggl \rceil \mathrm{for\ } p \in P.
\end{equation}
We then make the estimation for this representative layer configuration. Note that for all parameters which have a linear influence on the execution time behavior the step width is $w_p = 1$ and, therefore, $x_p = p$ for $p \in P$.

This approach is effective because for the observed step-wise execution time behavior of the accelerators in question, all configurations within the same step exhibit very similar execution times. 

\vspace{-4mm}
\subsubsection{Multi-Layer and Network Execution Time Estimation}
To be able to make execution time estimations for whole DNNs, we opted to take a similar approach to \cite{Wess2021} and \cite{Zhang2021}. The authors of ANNETTE use two different types of so called 'multi-layer benchmarks' to capture layer fusion, while Zhang et al. call a set of operators that are typically fused during model deployment an 'execution kernel'. When estimating whole DNNs in both papers, the authors search for those subgraphs in the complete model graph, make estimations for the subgraphs and sum the individual estimated execution times. 

We also use the most common building blocks of DNNs for which we want to make an estimation as multi-layer benchmarks. As those building blocks typically contain a type of Convolution (Conv2D, DepthwiseConv2D, PointwiseConv2D), we will again sample from the corresponding set of PRs to determine the parameters of the Convolution, select the resulting shapes of the remaining layers in the building block and carry out our benchmarks. These measurements are then used to determine how the runtime of a complete building-block behaves compared to the sum of individual execution times of the contained layers.

In the estimation phase for a whole DNN we, firstly, determine which building blocks are contained in the network and the layers and parameters inside the building block. Secondly, we map the configuration of the building block's convolutional layers to the corresponding PRs as described in Eq.~\ref{eq:pr_mapping}. And, thirdly, we use those PRs to make estimations for the contained layers. 

Depending on the platform, the estimation for a building block can differ. For the undisclosed commercial AI accelerator with two separate FUs which can execute certain layers in parallel, we estimate the execution time $t_b$ of a building block $b$ as follows:
\begin{equation}
\label{eq:undisclosed_bb_estimation}
    t_b = \begin{cases}
        \max\lbrace t_l \mathrm{\ for\ } l \in b \rbrace & 
        \begin{split}
            \mathrm{if\ } b \in \lbrace &PWConv2D(DWConv2D(x)),\\
            &FullyConnected(Pool(x)) \rbrace
        \end{split} \\
        \textstyle\sum_{l \in b}{t_l} &  \mathrm{else\ }
    \end{cases} \\
\end{equation}
Note that the undisclosed commercial AI accelerator fuses all ReLU layers with the preceding layers. Therefore, ReLU layers do not need to be considered in the sum of execution times.

In the case of the NVIDIA Jetson AGX Xavier GPU, we observed that the runtime of a building block is shorter than the sum of the individual runtimes of the contained layers. 
Therefore, we measured about 500 configurations per building-block to determine a so called fusing factor. The individually estimated layer execution times will be summed, corrected by this fusing factor, resulting in the execution time of a building block:%
\vspace{-2mm}
\begin{equation}
\label{eq:agx_bb_estimation}
    t_b = \textstyle\sum_{l \in b}{t_l} - f_\beta (b)
\end{equation}%
\vspace{-6mm}

\noindent The fusing factor $f_\beta (b)$ is calculated by a linear model for each type of building block $\beta$ (see multi-layer benchmarks in Sec.~\ref{subsec:benchmarks}) according to Eq.~\ref{eq:agx_bb_estimation_fusing_factor}. The weights $w_\beta$ and $c_\beta$ are fitted to the measured multi-layer benchmarks. $\#ops(b)$ denotes the total number of operations in a block $b$.%
\vspace{-2mm}
\begin{equation}
\label{eq:agx_bb_estimation_fusing_factor}
    f_\beta(b)= \#ops(b) \cdot w_\beta + c_\beta
\end{equation}%
\vspace{-6mm}

\noindent The execution time of the whole DNN is the sum of the contained building blocks' execution times:%
\vspace{-2mm}
\begin{equation}
\label{eq:whole_dnn_estimation}
    t_{DNN} = \textstyle\sum_{b \in B}{t_b}
\end{equation}

For DNN layers where we did not observe a step-wise execution time behavior, we sample up to \num{9000} data points and use these data points to train a Random Forest model for usage in the estimation of a building block's execution time.

%% file: 04_evaluation.tex
\section{Evaluation}
\label{sec:evaluation}
We evaluate our methodology for reducing the number of samples needed for training statistical estimation models for the UltraTrail accelerator (white-box), the VTA accelerator (gray-box), the undisclosed commercial AI accelerator (gray-box), and the NVIDIA Jetson AGX GPU (black-box).

\subsection{Experimental Setup}
We use the open-source tool ANNETTE\footnote{\url{https://github.com/embedded-machine-learning/annette}, last accessed 03/26/2024} for generating single-layer and multi-layer benchmarks, which we extended to use PyTorch instead of TensorFlow models. These models are then deployed to the different platforms using different backends. For UltraTrail we use the custom C-interface of the accelerator together with the RTL simulation. As VTA is part of the Apache TVM (version 0.15.0-dev0) project, we use TVM to deploy our models on the accelerator and the Verilator RTL simulation to determine the execution time. The measurements for the undisclosed commercial AI accelerator are performed using the timing simulator provided by the vendor. Lastly, for deployment on the NVIDIA Jetson AGX Xavier GPU running CUDA 11.4, we also use TVM with the TensorRT 8.5.2 backend and record the execution times reported by TVM.
    
\subsection{Benchmarks}
\label{subsec:benchmarks}
As the NVIDIA Jetson AGX Xavier GPU and the undisclosed commercial AI accelerator support a very broad range of operators, we ran \textbf{single-layer benchmarks} for the layers that are contained in the DNNs we use for the whole DNN evaluation. Namely, Conv2D, Depthwise Conv2D, Pointwise Conv2D, Average-Pooling, Max-Pooling, and Fully-Connected. For UltraTrail we ran the supported Conv1D and for VTA we benchmarked the two supported layers, Conv2D and Fully-Connected (via the general matrix multiplication). All benchmarks were sampled from the respective set of PRs.

As \textbf{multi-layer benchmarks} we selected recurring building-blocks of layers, which form the majority of our DNNs used for whole DNN evaluation.
\begin{itemize}
    \item Depthwise Convolution followed by Pointwise Convolution, each followed by ReLU (also known as Depthwise Separable Convolution), as used in MobileNet \cite{Mobilenets2017}: \vspace{-4mm}
        \begin{center}$ReLU(PWConv2D(ReLU(DWConv2D(x))))$\end{center}
    Note that for the Jetson AGX GPU the influence of the ReLU layer on the execution time was negligible. Therefore, no dedicated estimation models were needed for ReLU on this platform.
    \item A block of two Convolutions with a shortcut connection followed by an element-wise addition with ReLU, as used in ResNet \cite{ResNet2015}: 
        \begin{center}$ReLU(Add(x, Conv2D_1(ReLU(Conv2D_0(x)))))$\end{center}
    \item A block of two Convolutions with a shortcut connection that contains another Convolution, followed by an element-wise addition with ReLU, as used in ResNet \cite{ResNet2015}:
    \begin{center}$ReLU(Add(Conv2D_2(x), Conv2D_1(ReLU(Conv2D_0(x)))))$\end{center}
    \item A Max- or Average-Pooling layer followed by a Fully-Connected layer:
        \begin{center} $FullyConnected(Pool(x))$ \end{center}
    The execution time behavior of Max- and Average-Pooling was identical on the undisclosed commercial AI accelerator and the Jetson AGX GPU. Therefore, a combined Pooling model could be used. This multi-layer benchmark was only performed for the undisclosed commercial AI accelerator, in order to verify the parallel execution on the two separate FUs.
\end{itemize}

\subsection{Test Set and Sampling Strategy Comparison}
To show that the models we build from the PRs can actually estimate the layer execution times of state-of-the-art DNNs we chose the layers from the Keras model zoo\footnote{\url{https://keras.io/api/applications/}, last accessed 03/26/2024} to build our test set. These layer parameters have not been seen during training of our estimation models and will be used for evaluating the estimation accuracy. For the UltraTrail accelerator, we use the Conv1D layers of the TC-ResNet8 \cite{Choi2019} as our test set.

To compare our method for selecting benchmark points against an uninformed approach, we sampled different data set sizes randomly from the set of PRs and randomly from the complete parameter space. We then trained Random Forest estimation models using both sampling methods and evaluated the accuracy of the prediction using the test set. If not otherwise specified, we report the mean absolute percentage error (MAPE) of the estimation over all the test set layers. An overview of the results is depicted in Table~\ref{tab:best_mape} as well as the mean measurement time of a single benchmark point on the different platforms.
In case of the NVIDIA Jetson AGX Xavier GPU we performed each measurement \num{500} times and used the median in order to mitigate the influence of the first few slower runs, which were due to setup overhead. As the other measurements were performed using simulators, there was no need to perform multiple runs of the same benchmark point.

\begin{table}[htbp]
    \centering 
    {\scriptsize
    \input{table_accelerator_comparison}
    }
    \vspace{2mm}
    \caption{Overview of the dataset size and RMSPE for the best MAPE results for all AI accelerators and layer types for which the PR sampling and mapping was applied. The last column shows the mean measurement time of a single benchmark point for the different platforms. For the Jetson AGX we performed \num{500} runs for each benchmark.}
    \label{tab:best_mape}
\end{table}%
\vspace{-13mm}

\subsection{UltraTrail (white-box)}
\begin{figure}[htbp]
\centering
\includegraphics[width=\textwidth]{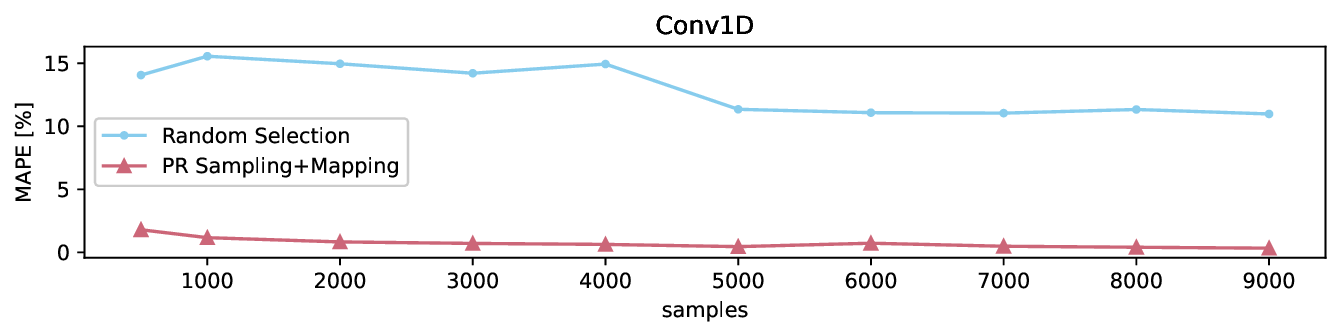}
\vspace{-8mm}
\caption{Comparison of the estimation accuracy when sampling from the set of PRs vs. random sampling from the complete parameter space for the Conv1D layer on the UltraTrail accelerator.}
\label{fig:evaluation_ultra_trail}
\end{figure}

Fig.~\ref{fig:evaluation_ultra_trail} shows how the estimation accuracy develops when using different amounts of training data. In this case, sampling a dataset of only \num{9000} layers from the set of PRs, yields an excellent MAPE of only \SI{0.33}{\percent}. This stems from the fact that the execution time behavior of the UltraTrail accelerator is extremely regular and the measurements of the RTL simulation do not contain any noise. Training the estimation model with the same amount of randomly sampled data points only reaches an accuracy of \SI{10.98}{\percent}. 

\noindent We do not present whole DNN estimations for UltraTrail because the execution time of multiple layers is the sum of the single layer execution times \cite{Bernardo2020}.

\subsection{VTA (gray-box)}
\begin{figure}[htbp]
\centering
\includegraphics[width=\textwidth]{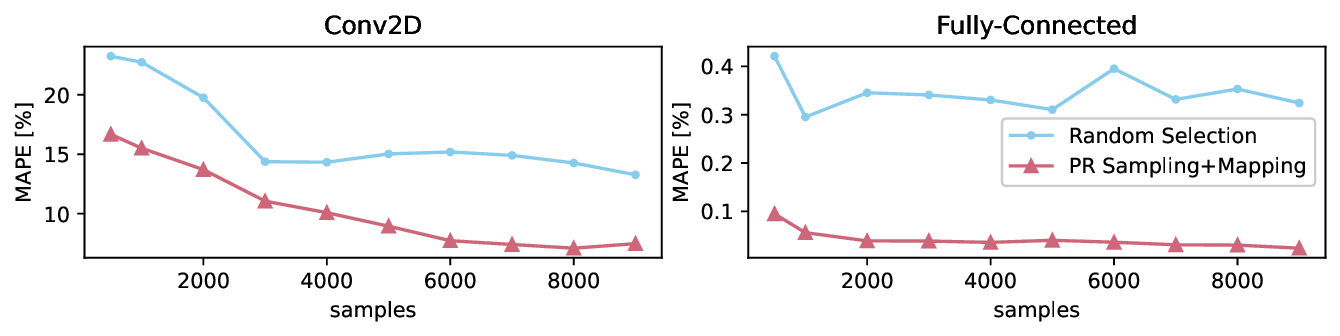}
\vspace{-8mm}
\caption{Comparison of the estimation accuracy when sampling from the set of PRs vs. random sampling from the complete parameter space for the Conv2D and Fully-Connected layer on the VTA accelerator.}
\vspace{-5mm}
\label{fig:evaluation_vta}
\end{figure}

Looking at the results of VTA, one of the gray-box accelerators, in Fig.~\ref{fig:evaluation_vta} one can clearly see that our approach achieves a lower MAPE for the execution time estimation of Conv2D and Fully-Connected layers compared to random sampling from the complete set of layer parameters. For this accelerator, we achieved our top result for the Conv2D layer of only \SI{7.09}{\percent} MAPE and the Fully-Connected layer of only \SI{0.02}{\percent} MAPE.%
\vspace{-6mm}
\vspace{-4mm}

\subsection{Undisclosed Commercial AI Accelerator (gray-box)}
\begin{figure}[htbp]
\centering
\includegraphics[width=\textwidth]{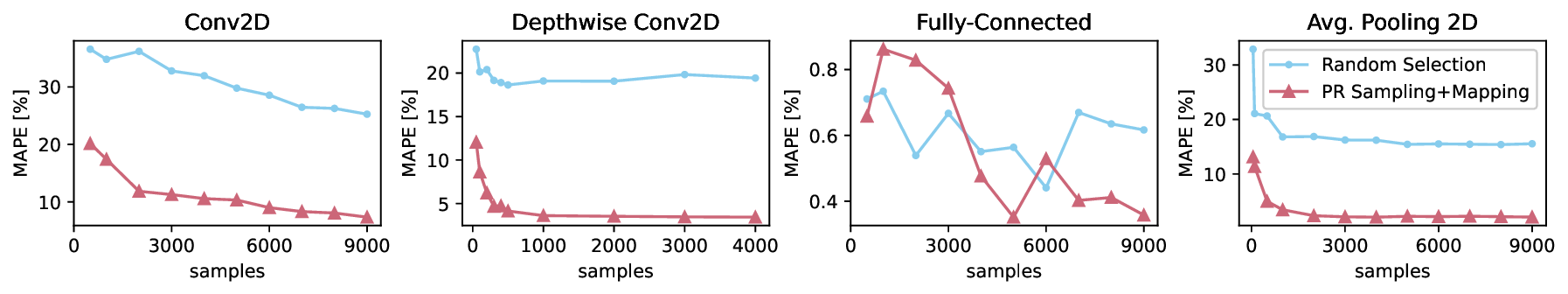}
\vspace{-8mm}
\caption{Comparison of the estimation accuracy when sampling from the set of PRs vs. random sampling from the complete parameter space for different layers on the undisclosed commercial AI accelerator.}
\vspace{-4mm}
\label{fig:evaluation_undisclosed}
\end{figure}

When comparing PR sampling and mapping against random sampling in Fig.~\ref{fig:evaluation_undisclosed}, one can again see that our method results in a better estimation accuracy overall, but additionally the MAPE decreases much faster when increasing the training dataset size. Only for the Fully-Connected layer, our method does not outperform the random selection. But as we can see, the MAPE is always below \SI{1}{\percent}, meaning that the execution time of the Fully-Connected layer is straightforward to estimate even with very few training samples. %
\vspace{-5mm}

We also applied our estimation models to estimate the execution time of whole DNNs using Eq.~\ref{eq:undisclosed_bb_estimation} and \ref{eq:whole_dnn_estimation} to account for overlapping layer execution due to the two distinct FUs of the undisclosed commercial AI accelerator. The results are displayed in Table~\ref{tab:whole_dnn_comparison}. We achieve an excellent estimation accuracy due to the step-wise execution time behavior of several layers on this accelerator as well as our knowledge about the parallel execution.%
\vspace{-4mm}
\begin{table}[]
    \centering 
    {\scriptsize
    \input{table_overview}
    }
    \vspace{2mm}
    \caption{Whole DNN estimation results for the undisclosed commercial AI accelerator and the NVIDIA Jetson AGX Xavier GPU.}
    \vspace{-6mm}
    \label{tab:whole_dnn_comparison}
\end{table}

\subsection{NVIDIA Jetson AGX Xavier GPU (black-box)}
The NVIDIA Jetson AGX Xavier GPU is not a highly specialized accelerator, but an embedded GPU. We still wanted to show that our approach can be used with this kind of computing device.

\begin{figure}[htbp]
\centering
\includegraphics[width=\textwidth]{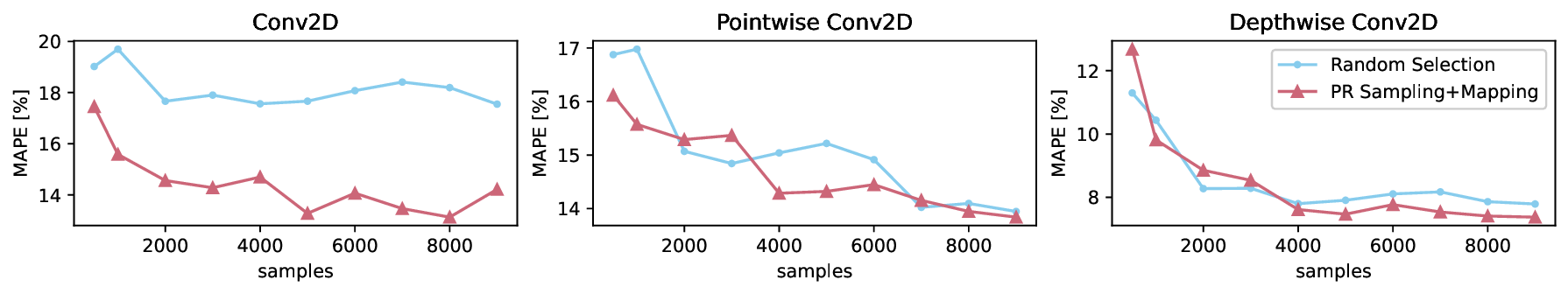}
\vspace{-8mm}
\caption{Comparison of the estimation accuracy when sampling from the set of PRs vs. random sampling from the complete parameter space for different layers on the NVIDIA Jetson AGX Xavier GPU.}
\vspace{-4mm}
\label{fig:evaluation_agx}
\end{figure}

In Fig.~\ref{fig:evaluation_agx} we see the MAPE of the single-layer estimations using PR sampling and mapping, and random sampling from the complete parameter space. The estimation accuracy is generally worse for the AGX Xavier GPU compared to the other AI accelerators, which can be explained by the fact that for this black-box accelerator we have almost no information to guide the sweep benchmarks or the selection of the PR. Moreover, the complexity of an embedded GPU is generally higher than for specialized accelerators. Still, for Conv2D our approach yields better results compared to random sampling from such a small training dataset. For the Pointwise Conv2D and the Depthwise Conv2D our approach does not outperform the random selection.

This can be attributed to the fact that only one parameter (input channels $C$) has a step-wise influence on the execution time behavior of the Depthwise Conv2D, resulting in a set of PRs that is similar to the complete set of layer configurations. For the Pointwise Conv2D this can be explained by the overall very short execution times of this layer type (longest observed execution time during parameter sweeps was \SI{0.28}{\milli\second}). For such fast layers, the overhead associated with the execution of the operations amounts to a bigger portion of the total execution time, while our approach focuses on the computational part.%
\vspace{-5mm}

The results for the whole DNN estimations can be seen in Table~\ref{tab:whole_dnn_comparison}. The estimation accuracy is worse compared to the undisclosed commercial AI accelerator. This can be attributed again to the fact that the Jetson AGX GPU is a black-box accelerator, for which we have almost no information, especially when it comes to how multiple layers are mapped onto the GPU architecture together.%
\vspace{-2mm}

\subsection{Comparison to State-of-the-Art}
A comparison to other state-of-the-art (SOTA) execution time estimation approaches is presented in Table~\ref{tab:sota_comparison}. If not otherwise stated, it shows the literature reported RMSPE and MAPE values of the approaches presented in~\ref{sec:related-work} Related Work. Note that often only one of the metrics is available. Our approach has comparable accuracy to the other SOTA methods, while needing much fewer training samples, especially compared to the best approach for estimating execution time on the NVIDIA Jetson AGX Xavier GPU by Bouzidi et al. \cite{Bouzidi2021} who used approximately \num{200000} whole DNN benchmarks for training their estimator.%

\vspace{-4mm}

\begin{table}[htb]
    \centering 
    {\scriptsize
    \input{table_estimator_comparison}
    }
    \vspace{2mm}
    \caption{Comparison with SOTA execution time estimation methods. \\
    $^{1}$values were computed from the reported times, 
    $^{2}$size of the dataset for the Conv+bn+relu execution kernel, 
    $^{3}$the authors reported RMSPE separately for each DNN, 
    $^{4}$per layer type, 
    $^{5}$per building-block for the fusing factor on the AGX}
    \label{tab:sota_comparison}
\end{table}

%% file: table_accelerator_comparison.tex
{
\setlength{\tabcolsep}{0.4em}
\begin{tabular}{ccccc|c}
    \toprule
    AI Accelerator & Layer Type Size & Dataset Size & RMSPE & MAPE & Mean Meas. Time [\si{\s}]\\
    \midrule
    UltraTrail & Conv1D & 9000 & \SI{0.51}{\percent} & \SI{0.33}{\percent} & \SI{0.38}{} ($SD=0.07$)\\
    \midrule
    \multirow{2}*{VTA} 
    & Conv2D & 8000 & \SI{4.49}{\percent} & \SI{7.09}{\percent} & \SI{220.74}{} ($SD=549.60$)\\
    & Fully-Connected & 9000 & \SI{0.44}{\percent} & \SI{0.02}{\percent} & \SI{2.21}{} ($SD=1.65$)\\
    \midrule
    \multirow{4}*{\shortstack{Undisclosed\\Commercial\\AI Accelerator}} 
    & Conv2D & 9000 & \SI{9.93}{\percent} & \SI{7.35}{\percent} & \SI{5.34}{} ($SD=0.36$)\\
    & Depthwise Conv2D & 4000 & \SI{11.59}{\percent} & \SI{3.44}{\percent} & \SI{5.29}{} ($SD=0.22$)\\
    & Fully-Connected & 5000 & \SI{0.34}{\percent} & \SI{0.35}{\percent} & \SI{5.36}{} ($SD=0.12$)\\
    & 2D Average Pooling & 4000 & \SI{6.09}{\percent} & \SI{2.09}{\percent} & \SI{5.29}{} ($SD=0.05$)\\
    \midrule
    \multirow{3}*{\shortstack{NVIDIA Jetson\\AGX GPU}} 
    & Conv2D & 8000 & \SI{27.06}{\percent} & \SI{13.13}{\percent} & \SI{47.59}{} ($SD=13.29$)\\
    & Pointwise Conv2D & 9000 & \SI{29.99}{\percent} & \SI{13.84}{\percent} & \SI{46.00}{} ($SD=11.22$)\\
    & Depthwise Conv2D & 9000 & \SI{11.43}{\percent} & \SI{7.37}{\percent} & \SI{65.12}{} ($SD=40.65$)\\
    \bottomrule
\end{tabular}
}

%% file: table_overview.tex
{
\setlength{\tabcolsep}{0.4em}
\begin{tabular}{ccccc}
    \toprule
    AI Accelerator & DNN & Meas. time [\si{\milli\second}] & Est. time [\si{\milli\second}] & \% Error\\
    \midrule
    \multirow{2}*{\shortstack{Undisclosed Commercial\\AI Accelerator}}     & MobileNet \cite{Mobilenets2017} & \num{2.95}    & \num{2.93}    & \SI{0.68}{\percent}\\
                            & ResNet18 \cite{ResNet2015}  & \num{4.88}    & \num{4.63}    & \SI{5.12}{\percent}\\
    \midrule
    \multirow{2}*{\shortstack{NVIDIA Jetson\\AGX GPU}}    & MobileNet \cite{Mobilenets2017} & \num{1.79}    & \num{2.14}    & \SI{19.55}{\percent}\\
                                            & ResNet18 \cite{ResNet2015}  & \num{2.90}    & \num{3.47}    & \SI{19.66}{\percent}\\
    \bottomrule
\end{tabular}
}

%% file: table_estimator_comparison.tex
{
\setlength{\tabcolsep}{0.4em}
\begin{tabular}{cccccc}
    \toprule
    Work & Type & Platform & Dataset Size & RMSPE & MAPE\\
    \midrule
    \multirow{4}*{ANNETTE \cite{Wess2021}} & \multirow{2}*{Conv2D Layer} & NCS2 & \multirow{2}*{\shortstack{\num{35000}\\ per platform}} & \SI{42.60}{\percent} & \SI{15.57}{\percent}\\
    & & ZCU102 & & \SI{10.55}{\percent} & \SI{12.71}{\percent} \\
    \cmidrule(lr){2-6}
    & \multirow{2}*{Whole DNN} & NCS2 & \num{36570} & - & \SI{7.44}{\percent} \\
    & & ZCU102 & \num{37812} & - & \SI{3.47}{\percent} \\
    \midrule
    \multirow{4}*{Blackthorn \cite{Lechner2021}} & \multirow{2}*{Conv2D Layer} & Jetson Nano & \multirow{2}*{\shortstack{\num{15000}\\ per platform}} & \SI{5.89}{\percent} & - \\
    & & Jetson TX2 & & \SI{6.10}{\percent} & - \\
    \cmidrule(lr){2-6}
    & \multirow{2}*{Whole DNN} & Jetson Nano & \multirow{2}*{\shortstack{no multi-layer\\models}} & \SI{1.71}{\percent}$^{1}$ & \SI{2.95}{\percent}$^{1}$ \\
    & & Jetson TX2 & & \SI{3.06}{\percent}$^{1}$ & \SI{4.29}{\percent}$^{1}$ \\
    \midrule
    \multirow{2}*{Bouzidi et al. \cite{Bouzidi2021}} & \multirow{2}*{Whole DNN} & Jetson AGX & \multirow{2}*{\shortstack{$\sim$\num{200 000} DNNs\\per platform}} & - & \SI{7.67}{\percent}\\
    & & Jetson TX2 & & - & \SI{8.37}{\percent}\\
    \midrule
    \multirow{3}*{nn-Meter \cite{Zhang2021}} & \multirow{3}*{Whole DNN} & Cortex-A76 & \num{15824}$^{2}$ & \qtyrange[range-phrase=~--~]{2.76}{5.54}{\percent}$^{3}$ & - \\
    & & Adreno 640 & \num{14040}$^{2}$ & \qtyrange[range-phrase=~--~]{1.35}{5.32}{\percent}$^{3}$ & - \\
    & & NCS2 & \num{39968}$^{2}$ & \qtyrange[range-phrase=~--~]{4.26}{22.25}{\percent}$^{3}$ & - \\
    \midrule
    \multirow{4}*{This} & \multirow{2}*{Conv2D Layer} & Undisclosed & 9000 & \SI{9.93}{\percent} & \SI{7.35}{\percent}\\
    & & Jetson AGX & 8000 & \SI{27.06}{\percent} & \SI{13.13}{\percent}\\
    \cmidrule(lr){2-6}
    % & \multirow{2}*{Whole DNN} & Undisclosed & \multirow{2}*{max. \num{9000}} & \SI{4.53}{\percent} & \SI{2.90}{\percent}\\
    % & & Jetson AGX & & \SI{20.17}{\percent} & \SI{19,60}{\percent}\\
    & \multirow{2}*{Whole DNN} & Undisclosed & \multirow{2}*{\shortstack{max. \num{9000}$^{4}$\\ + max. \num{500}$^{5}$}}  & \SI{4.53}{\percent} & \SI{2.90}{\percent}\\
    & & Jetson AGX &  & \SI{20.17}{\percent} & \SI{19,60}{\percent}\\
    
    \bottomrule
\end{tabular}
}

%% file: 99_conclusion.tex
\section{Conclusion and Future Work}
\label{sec:conclusion}
This paper presented a performance modeling methodology and associated benchmarking strategy to reduce the number of training samples for statistical performance models, and thus reduce the required measurement time while maintaining comparable estimation accuracy to state-of-the-art estimators. We leverage knowledge about the AI accelerator architecture and mapping to determine Performance Representatives (PR) algorithmically for the benchmarking and, later, estimation phase. We evaluated our approach for four different AI accelerators and, depending on the available hardware architecture knowledge, achieved an execution time estimation MAPE between \SIrange{0.02}{13.84}{\percent} for single-layers and \SIrange{2.90}{19.60}{\percent} for whole DNNs.

As future work, we plan to use the PRs of an AI accelerator as a search space constraint in hardware-aware NAS and to extend the PRs to additionally include the memory characteristics of the AI accelerator.